# Are Near Earth Objects the Key to Optimization Theory?

**Richard A. Formato**[1]

*Abstract:* This note suggests that near earth objects and Central Force Optimization have something in common, that NEO theory may hold the key to solving some vexing problems in deterministic optimization: local trapping and proof of convergence. CFO analogizes Newton's laws to locate the global maxima of a function. The NEO-CFO nexus is the striking similarity between CFO's $D_{avg}$ and an NEO's $\Delta V$ curves. Both exhibit oscillatory plateau-like regions connected by jumps, suggesting that CFO's metaphorical "gravity" indeed behaves like real gravity, thereby connecting NEOs and CFO and being the basis for speculating that NEO theory may address difficult issues in optimization.

*7 December 2009*
*Brewster, Massachusetts*

# Are Near Earth Objects the Key to Optimization Theory?

**Richard A. Formato[1]**

This note suggests that the theory of gravitationally trapped near earth objects (NEOs) provides an analytical framework for the further theoretical development of Central Force Optimization (CFO) (1-4). NEO theory may lead to deterministic mitigation of local trapping (a significant problem for many optimization algorithms). It also may lead to a proof of convergence (a milestone achievement for any algorithm). Applying NEO theory likely requires collaboration between theorists in celestial mechanics and optimization. Hopefully these observations will stimulate that collaboration.

CFO locates the global maxima ("fitnesses") of an "objective function" $f(x_1, x_2, ..., x_N)$ defined on an *N*-dimensional (*n*-D) "decision space" (DS) with unknown topology. CFO is a Nature-inspired metaheuristic like Particle Swarm Optimization and Ant Colony Optimization. But unlike PSO and ACO, it is deterministic instead of stochastic. CFO analogizes gravitational kinematics, thus embracing the metaphor of Newton's precise laws of gravity and motion.

The NEO-CFO connection is illustrated using the *n*-D step function in 2D, a recognized optimization benchmark that can be visualized. Its definition is
$$f(x) = -\sum_{i=1}^{N} \left( \lfloor x_i - x_o^i + 0.5 \rfloor \right)^2, \quad -100 \leq x_i \leq 100$$
(here $N = 2$, $x_o^1 = 75$, $x_o^2 = 30$). The highly discontinuous step is unimodal with a maximum of zero at $(75, 30)$. Figs. 1(a) and (b) plot it over its domain and in the vicinity of the maximum.

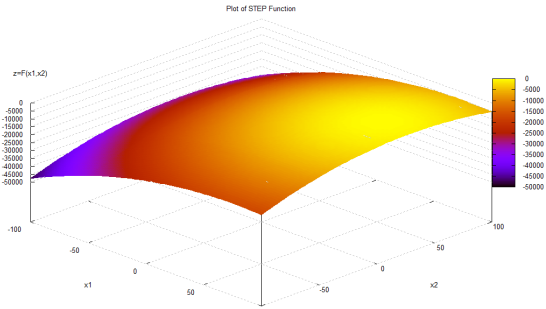 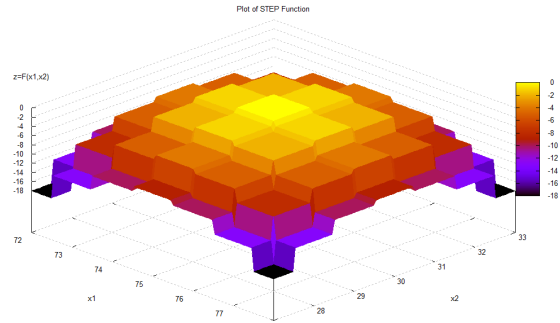

Fig. 1(a)   Fig. 1(b)

CFO flies "probes" through DS over "time steps" (iterations). Their trajectories are computed from two *equations of motion* analogous to the equations of motion for masses moving through space under the influence of real gravity. CFO "mass" is created by defining a function of the objective function's fitness.

The equations of motion for the probes' acceleration and position vectors are
$$\vec{a}_{j-1}^p = G \sum_{\substack{k=1 \\ k \neq p}}^{N_p} U(M_{j-1}^k - M_{j-1}^p) \cdot (M_{j-1}^k - M_{j-1}^p)^\alpha \times \frac{(\vec{R}_{j-1}^k - \vec{R}_{j-1}^p)}{\left\| \vec{R}_{j-1}^k - \vec{R}_{j-1}^p \right\|^\beta} \quad \text{and} \quad \vec{R}_j^p = \vec{R}_{j-1}^p + \frac{1}{2} \vec{a}_{j-1}^p \Delta t^2, \quad j \geq 1,$$
$M_{j-1}^p = f(x_1^{p,j-1}, x_2^{p,j-1}, ..., x_N^{p,j-1})$. Indices $1 \leq p \leq N_p$ and $0 \leq j \leq N_t$ are the probe and

iteration numbers, $N_p$ and $N_t$ being the total numbers. $U(\cdot)$ is the Unit Step,
$U(z) = \begin{cases} 1, & z \geq 0 \\ 0, & otherwise \end{cases}$, and $MASS_{CFO} = U(M_{j-1}^k - M_{j-1}^p) \cdot (M_{j-1}^k - M_{j-1}^p)^\alpha$.

Fig. 1(c) plots the trajectories of the probes with the best fitnesses, and Fig. 1(d) the individual probe trajectories. These plots are visually chaotic, providing no hint whatsoever of the underlying mathematical regularity that forms the NEO-CFO nexus.

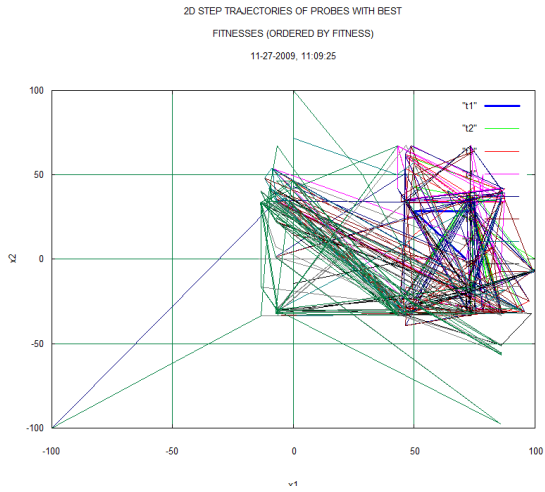
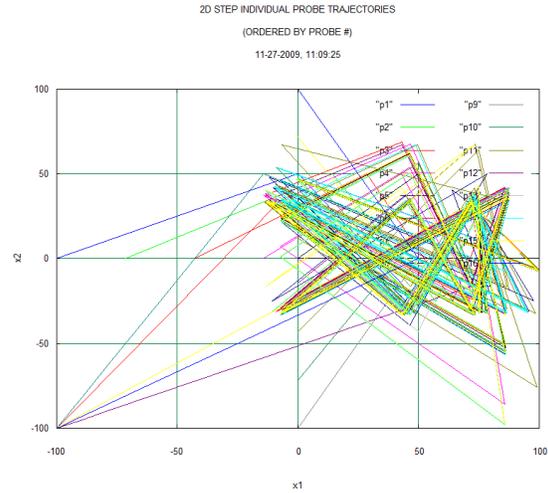

Fig. 1(c)  Fig. 1(d)

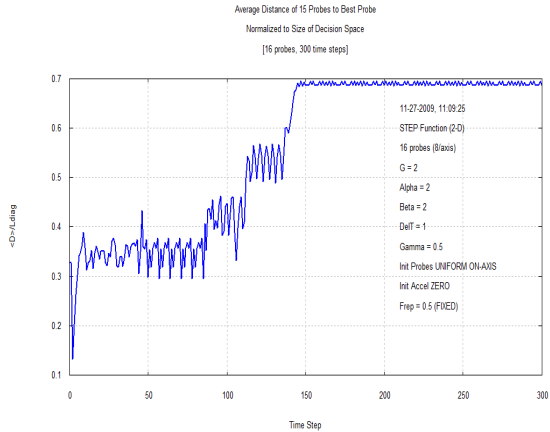
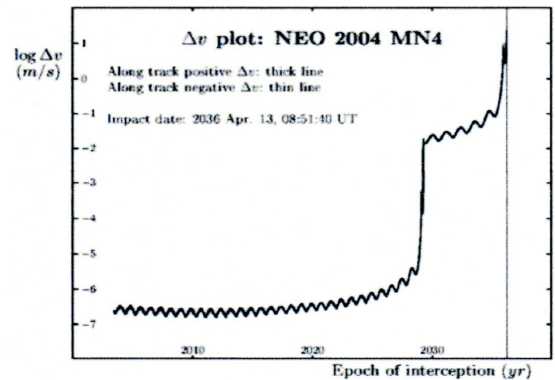

Fig. 1(e)  Fig. 1(f)

That regularity appears in CFO's $D_{avg}$ curve (normalized average distance between the probe with the best fitness and all others) plotted in Fig. 1(e) (annotated with run parameters). $D_{avg}$ exhibits four oscillatory plateaus connected by jumps. Although the oscillation may not be precisely repetitive, in many cases it is (starting at step 162, for example, $D_{avg}$ comprises the repeating sequence 0.6859416, 0.6917107, 0.6868708, 0.6952526, 0.6855014, 0.6956451, 0.6859433, 0.6917887, 0.6868326, 0.6859393, 0.6877515, 0.6939823, 0.6870971, 0.6956298, 0.6859431, 0.6866136, 0.6872625, 0.6940363, 0.6861267, 0.6953240).

Oscillation in $D_{avg}$ appears to be a reliable signal of local trapping (empirically determined). In this case, CFO is trapped at a local maximum of $-1$ at $(75, 28.57142857)$. Trapping caused CFO to miss the global maximum, often a problem with deterministic algorithms.

The connection between CFO and NEO theory is the structural similarity of $D_{avg}$ under trapping and the $\Delta V$ curve for a gravitationally trapped NEO. Fig. 1(f) plots asteroid Apophis' $\Delta V$ curve (reproduced from (5) with permission) computed by Professors Andrea Milani and Andrea Caruso using the theory of resonant returns (6) (private communication, Astronaut "Rusty" Schweickart). $\Delta V$ is the velocity change needed to avoid earth impact. $D_{avg}$ is a similar variable because it is proportional to velocity if $\Delta t$ is constant.

Apophis' $\Delta V$ curve contains two well-defined oscillatory plateaus connected by a jump and what appears to be the beginning of a third plateau, also connected by a jump, that is cut off by the vertical line marking earth impact in year 2036. The structural similarity to $D_{avg}$ is striking. Both curves comprise oscillatory plateaus connected by jumps, and it is difficult to imagine that their similarity is accidental. Rather, because the Apophis plot is based on real gravity trapping the asteroid in earth orbit, and $D_{avg}$ is based on CFO's metaphorical gravity trapping a probe at a local maximum, it is reasonable to believe that the similarity actually may be inevitable. If so, this observation is a compelling validation of the CFO gravitational metaphor, and the basis for speculating that NEO theory well may hold the key to solving important problems in optimization. Hopefully researchers with appropriate skills and interests will find out if it does.

**References and Notes**

1. R. A. Formato, *Prog. Electromagnetics Research*, PIER 77, 425-491 (2007): http://ceta.mit.edu/PIER/pier.php?volume=77 (DOI:10.2528/PIER07082403).
2. R. A. Formato, in *Nature Inspired Cooperative Strategies for Optimization (NICSO 2007),* **Studies in Computational Intelligence 129** (N. Krasnogor, *et al*., Eds.), vol. 129, Springer-Verlag, Heidelberg (2008).
3. R. A. Formato, *Int. J. Bio-Inspired Computation*, 1(4), 217-238 (2009). (DOI: 10.1504/IJBIC.2009.024721).
4. R. A. Formato, *OPSEARCH*, *Jour. of the Operations Research Society of India*, 46(1), 25-51 (2009). (DOI: 10.1007/s12597-009-0003-4).
5. R. Schweickart, *et al*.," arXiv:physics/0608155v1 (2006). http://arxiv.org/abs/physics/0608155.
6. G. B. Valsecchi, *et al*., Astronomy & Astrophysics, 408 (3), 1179-1196, 2003. http://www.aanda.org. (DOI: 10.1051/0004-6361:20031039).

[1]Richard A. Formato, JD, PhD
Registered Patent Attorney & Consulting Engineer
P.O. Box 1714, Harwich, MA 02645 USA
rf2@ieee.org